# Cobalt substitution induced magnetodielectric enhancement in multiferroic $Bi_2Fe_4O_9$

S. R. Mohapatra[1], P. N. Vishwakarma[1], S. D. Kaushik[2], R. J. Choudhary[3], N. Mohapatra[4], A. K. Singh[1*]

[1]Department of Physics and Astronomy, National Institute of Technology Rourkela-769008, Odisha, India.
[2]UGC-DAE Consortium for Scientific Research Mumbai Centre, R-5 Shed, BARC, Mumbai-400085, India.
[3]UGC-DAE-Consortium for Scientific Research Indore Centre, Indore 452 017, India.
[4]School of Basic Sciences, Indian Institute of Technology Bhubaneswar-171013, Odisha, India.

**Abstract:** Antiferromagnetic $Bi_2Fe_4O_9$ (BFO), lightly substituted by cobalt is studied for magnetodielectricity. The substitution causes a substantial decrease in the Néel temperature ($T_N$) from 250 K (in parent sample, BFO) to 152 K (in 2% Co substituted sample). At the same time, the substituted samples display a pronounced irreversibility in the ZFC-FC magnetization data for T < 370 K and opening of hysteresis in the M-H plot, thus signifying the onset of weak ferromagnetism (FM) and magnetic glassiness. The induced magnetic glassiness is found to slow down the dynamics such that the magnetization decay follows $M(t) \propto \exp[-(t)^{1-p}]$. The dielectric measurement in the same temperature window shows unusual oppression in $\varepsilon'$, for T~$T_N$ and contrasting nature of tan loss for temperatures above and below $T_N$, thus hinting a plausible coupling between the magnetic and electric order parameters. A confirmation to this coupling is seen in the magnetodielectric (MD) results, in which it is found that the substitution induces an additional component in the MD, apart from the usual components in BFO. This additional component of MD is found to obey $\propto \exp(\omega)^n$ behaviour, with the 'n' values being comparable to '1-p' of magnetization. The temperature variation of MD also shows a contrasting behaviour for the parent and 2% Co substituted sample with an enhancement of two times in MD value. In summary, our study shows ME coupling introduced by the magnetic glassiness and its behaviour is very much different from the intrinsic one.

**Keywords:** Magnetic glassiness; Magnetodielectric effect; Antiferromagnetism.

# I. INTRODUCTION

Multiferroic materials displaying intriguing magnetoelectric (ME) effect near room temperature are in focus of current interest owing to its rich potential technological influence along with noteworthy multifunctional device applications [1-5]. The ME coupling offers additional degrees of freedom, where electric as well as magnetic polarization can be tailored as it is sensitive to both the magnetic and electric fields, respectively [6]. However, near room temperature ME materials prevailing such striking features are scarcely found due to its mutually exclusive characteristics [2]. Still, BiFeO$_3$ emerges to be a leading candidate of room-temperature multiferroic (ferroelectricity due to Bi-lone pair and magnetism due to magnetic Fe$^{3+}$ ions) despite of weak coupling of the order parameters [7]. Similarly, hexagonal $R$MnO$_3$ ($R$ = rare earth elements) also shows relatively weak coupling and the cause for multiferroicity here is tilting of MnO$_5$ polyhedra [8]. Conversely, the search for pronounced coupling led to the discovery of few improper magnetic ferroelectrics where ferroelectricity arises due to lattice inversion symmetry breaking by modulated magnetic order, and the typical examples are orthorhombic $R$MnO$_3$, $R$Mn$_2$O$_5$ ($R$ = Tb, Ho, Dy, Y and Bi), magnetically frustrated Ni$_3$V$_2$O$_8$ etc [9-11]. Generally, most of the ME multiferroics found till date are functional and applicable at cryogenic temperatures.

Recently, the ME coupling has been explored in partially disordered La$_2$NiMnO$_6$ near room temperature [12], multi-glass system (Sr,Mn)TiO$_3$ [13], disordered multiferroics such as Sr$_{0.98}$Mn$_{0.02}$TiO$_3$, EuTiO$_3$, Pb(Fe$_{0.5}$Nb$_{0.5}$)O$_3$ and CuCr$_{1-x}$In$_x$P$_2$S$_6$ [14], spin-cluster-glass Fe$_2$TiO$_5$ [15] etc. While ME coupling is broadly studied among various magnetically ordered materials, now a days a lot of investigations are focussed on to various spin glass or disordered systems as well [14, 15]. The challenge is to gain a profound understanding of the rich magneto(di)electric phenomena and widen the scope to put them to pragmatic use.

The compound of our interest is mullite-type Bi$_2$Fe$_4$O$_9$ [16, 17], a material prototype of the Cairo spin lattice [18] which is a well-known antiferromagnetically (AFM) ordered system [19, 20]. Bi$_2$Fe$_4$O$_9$ (BFO) crystallizes in orthorhombic space group '*Pbam*' and its crystal structure was first reported by Niizeki et al. [21]. BFO shows many interesting properties such as catalytic [22], gas sensors [23], electronic [24], magnetic [25], crystal chemical [26, 16] and temperature-dependant behaviours [27-29]. BFO is characterised by the linear chains of edge-sharing FeO$_6$ octahedra running parallel to *c*-axis. These chains are further linked together by corner-sharing Fe$_2$O$_7$ double tetrahedra units and BiO$_6$E groups (where E refers to 6s$^2$ lone electron pair) along the *c*-axis in an ordered and alternate fashion. The corner-sharing tetrahedra are coupled perpendicular to the edge sharing FeO$_6$ octahedral chains, thus linking the chains to each other [27, 16]. BFO shows a peculiar pentagon spin frustration due to competing exchange interactions of different types of $Fe^{3+}$ ions leading to noncollinear magnetic structure [30]. In fact, the real geometry of BFO differs from the Cairo pentagonal lattice. From Fig. 1 (a) it can be seen that BFO comprises of two stacked octahedral coordinated Fe (O) i.e. *β1* & *β2* atoms at two vertices in each pentagon instead of a single. The structure shows, five main magnetic super-exchange interactions i.e. $J_1$ to $J_5$ ($J_1$ & $J_2$ is not shown here). Among them, $J_4$ (i.e. interaction between tetrahedral $Fe^{3+}$ ions denoted by α1 & α2) is the strongest and AFM in nature followed by $J_5$ and $J_3$. The projection along c-axis of this layer forms a pentagonal lattice having three slightly different bonds per pentagon. Previous reports on structural and magnetic properties by Giaquinta et al. [31, 32] stated that diamagnetic substitution (Ga and Al) at Fe-site caused spin frustration and revealed spin-glass like behaviour with decrease in the Néel temperature ($T_N$). In recent past, several researchers have tried to improve the multiferroic properties by chemically substituting various dopants at Bi and Fe-site of BFO, thus causing distortion/disorderness in the system [33-36].

In this paper, we report the substitutional effect of cobalt ions at Fe-site in context to dielectric, magnetic and magnetodielectric aspect below room temperature. Being a transition metal $Co^{3+}$ is chosen for substitution at $Fe^{3+}$ site because (a) $Co^{3+}$ (61 pm for high spin state (HS) & 55 pm for low spin (LS) state under coordination no. = 6) and $Fe^{3+}$ (65 pm for HS state, 55 pm for LS state under coordination no. = 6 and 49 pm for HS state under coordination no. = 4) possesses similar ionic radii [37], (b) the AFM exchange interaction between $^5dFe^{3+}$ (5.92 $\mu_B$ per $^5dFe^{3+}$) and $^6dCo^{3+}$ (4.9 $\mu_B$ per $^6dCo^{3+}$) would induce slight disorderness in the substituted samples which may result in enhanced properties. Here, we report the effect of light cobalt substitution on the magnetic, dielectric and magnetodielectric properties of multiferroic BFO.

## II. EXPERIMENTAL

The polycrystalline samples of $Bi_2(Fe_{1-x}Co_x)_4O_9$ (x = 0, 0.005, 0.01, 0.015 and 0.02) were prepared by conventional solid state reaction route using high purity oxides (99.9%, Sigma Aldrich) namely $Bi_2O_3$, $Fe_2O_3$ and $Co_3O_4$. The above chemicals were thoroughly mixed in stoichiometric proportions, grounded for 2 h using an agate mortar and calcined at 1073 K for 12 h. The calcined powders were again grounded for 2 h and then compacted into cylindrical pallets of 10 mm diameter followed by sintering at 1123 K for 10 h. The phase formation of the synthesized samples were verified by room-temperature X-ray diffraction (XRD) measurements carried out using a multipurpose X-ray diffraction system (RIGAKU, JAPAN). XRD data were collected using a Bragg–Brentano geometry equipped with a secondary Ni filter, Cu-Kα radiation (λ = 0.15406 nm) in a wide range of Bragg angles (10° ≤ 2θ ≤ 70°) with a step size of 0.002° at a scan rate of 3°/min. Subsequently, the XRD data were analysed by the Rietveld refinement technique using FULLPROF package [38]. The X-ray photoemission spectroscopy (XPS) measurement was performed using the Omicron energy analyzer (EA 125) instrument with an Al-Kα (1486.6 eV) X-ray source. The pressure of the analyzer chamber was maintained in the order of $1.33 \times 10^{-8}$ Pa during the XPS measurement. The spectrometer was calibrated using Au-$4f_{7/2}$ core level (at binding energy of 84 eV) spectra. The values corresponding to C 1s peak were used as a reference for spectrum analysis. Magnetization measurements were carried out in the 9T re-liquefier based Physical Property Measurement System (PPMS)-Vibrating Sample Magnetometer (Quantum Design). The dielectric measurements were performed using high precession impedance analyser (Wayne Kerr 6500B) over a wide frequency range of 100 Hz − 100 kHz. For dielectric and DC resistance measurements contacts were made by applying silver paste on both the faces of all samples. The temperature variation for the dielectric measurement from 75 - 310 K was attained using a closed cycle refrigerator (Cryo Industries, USA). Lastly, the temperature dependent DC resistance measurement (200 - 305 K) was done using an electrometer (Keithley 6517B). The magnetic field (up to 1.3 T) required for the magnetodielectric and magnetoresistance measurement was achieved from an electromagnet (GMW 5403) equipped with bipolar DC power supply.

## III. RESULTS AND DISCUSSION

### A. X-ray diffraction (XRD) and X-ray photoelectron spectroscopy (XPS) study:

Fig. 1 (b) shows the Rietveld refined room-temperature powder X-ray diffraction (XRD) patterns of $Bi_2(Fe_{1-x}Co_x)_4O_9$ (0 ≤ x ≤ 0.02) samples abbreviated as BFO (x = 0), BFCO0.5 (x = 0.005), BFCO1 (x = 0.01), BFCO1.5 (x = 0.015) and BFCO2 (x = 0.02), respectively. All the refinements of the atomic position parameters are found to be in good agreement with the previous reported values [21]. XRD pattern of x = 0, 0.005 indicates single phase formation (orthorhombic phase, 'Pbam' space group) whereas for x = 0.01 - 0.02 indicates the presence of traces of the perovskite type $BiFeO_3$ phase (rhombohedral phase, 'R3c' space group) which is < 4 wt. % for x = 0.01 - 0.02. The fitted parameters obtained from the refinements are listed in table 1, the corresponding Rietveld plots given in Figure 1 indicate a good agreement between the observed and refined patterns. The refined XRD data mentioned in table 1 show a decrease in the lattice parameters as well as in the cell volume with increasing

Co content. This decrease is attributed to the slightly lower ionic radius of $Co^{3+}$ substituted ions compared to that of $Fe^{3+}$ ions and thus confirms the substitution of cobalt ions at the Fe-site. Moreover, the substitutional effect on crystallite size as well as on micro-strain analysis determined using the Williamson and Hall method [39] are given in table 1. We observed a decrease in crystallite size i.e. from 247(9) nm for BFO to 118(3) nm for BFCO2 and an increase in micro-strain i.e. from $2.20(7) \times 10^{-3}$ % for BFO to $8.36(7) \times 10^{-3}$ % for BFCO2. Additionally, the interatomic metal-oxygen distance and positional parameters obtained from Rietveld refined XRD data of BFO and the highest substituted sample (BFCO2) are listed in table 2 and table 3, respectively. The XRD analysis shows that the substituted cobalt ions have slightly more affinity towards tetrahedral position than octahedral position (shown in table 3).

X-ray photoelectron spectroscopy (XPS) measurements are done to ascertain the oxidation states of Fe (Co) ions as high-temperature sintering may cause oxygen deficiency, thereby affecting the oxidation states of metal cations. Fig. 2(a, b) shows the XPS plot of BFO and BFCO2 sample in the energy range of Fe-2p core level spectra. Two broad peaks ~710 eV and ~723 eV, along with a broad hump in between and the characteristic satellite peaks separated by 13 eV on energy scale are seen. The asymmetrically broadening of the peaks suggests presence of one or more peaks in it. Thus, the peaks are deconvoluted for peaks corresponding to Fe $2p_{3/2}$ (709 eV & 711 eV) and Fe $2p_{1/2}$ (722.5 eV & 724.5 eV) [40, 41]. The deconvoluted peaks are found to be positioned at 709 eV & 710 eV and 722 eV & 724 eV, respectively. Presence of 709 eV and 722 eV peaks are associated with $Fe^{2+}$ and their presence in the spectra along with $Fe^{3+}$ peaks (710 eV and 724 eV), shows mixed nature of oxidation states of Fe in the sample. In the similar manner, spectra of Co ions partially substituted for Fe also suggests it to be in mixed oxidation states as seen from earlier report [42]. Integration of area under the respective peaks shows that the $Fe^{3+}:Fe^{2+}$ ratio in parent BFO sample is 80:20. To see the effect of cobalt substitution (if any) on the distribution of oxidation states, following the same procedure, the ratio 3+ : 2+ ions i.e., ($Fe^{3+}$ + $Co^{3+}$) : ($Fe^{2+}$ + $Co^{2+}$) in BFCO2 sample is evaluated. The evaluated ratio is found to remain unchanged i.e., 80:20, within the experimental errors. But it is to be noted that XPS being a surface sensitive method, the exact ratio of 3+ : 2+ ions is assumed to vary in bulk from above calculated values.

**B. DC magnetization:**

Fig. 3 (a-c) shows the temperature dependence of DC magnetic susceptibility ($\chi$ = M/H) of BFO, BFCO1 and BFCO2 samples under zero field cooled (ZFC) and field cooled (FC) conditions. The measurements are carried out in the temperature range of 75 - 380 K in an external field of 1 kOe. A noticeable broad cusp is seen around 250 K in the parent BFO sample, typically revealing its AFM behaviour and is consistent with earlier reports [31, 36]. To exactly locate the AFM transition temperature, the $d\chi/dT$ versus T (upper inset of Fig. 3 (a)) is plotted which confirms the Néel temperature ($T_N$) at 250 K. BFO is well known to be a magnetically frustrated system where the competing exchange interactions between the octahedral and tetrahedral coordinated Fe ions give rise to a unique pentagon frustration (shown in Fig. 1 (a)). The degree of frustration (*f*) for a spin frustrated system is defined as *f* = $|\theta_{CW}|/T_N$, where $|\theta_{CW}|$ is the Curie-Weiss temperature [43]. From the plot of inverse susceptibility ($1/\chi$) vs T (lower inset of Fig. 3 (a)), the value of $\theta_{CW}$ is determined to be - 2258 K, resulting the frustration parameter (*f*) to be 9.03. The obtained value of frustration parameter reveals highly frustrated AFM spin in BFO. It must be noted that though the system is highly frustrated, it is not magnetically disordered. This is revealed from the exactly superimposing ZFC-FC curves. On the other hand, the cobalt substituted samples display a large bifurcation in the ZFC-FC curves at low temperature and the bifurcation does not merge until temperature reaches ($T_{irr}$) ~370 K or above as shown in Fig. 3 (b, c). As the paramagnetic region of substituted samples is beyond the present temperature window of measurement, it is difficult to comment on the substitution induced effect on the frustration parameter. In the cobalt substituted samples as well, the $T_N$ is seen as an inflection in the ZFC plot of BFCO1 and BFCO2 at 158 ± 0.5 K and 152 ± 0.5 K, respectively. This is further confirmed from $d\chi/dT$ versus T plot (inset of respective plots). Fig. 3 (d) shows the variation of $T_N$ as the cobalt content (*x*) varies in the sample. It is to be noted that the absolute values of ZFC $\chi$ are about two times higher for BFCO1 and about seven times higher for BFCO2 than that of the parent sample (BFO). This is an indication of increased net magnetization due to weakening of AFM ordering. The FC data first decreases and then increases at low temperature, a behaviour distinctly different from the ZFC data which shows continuous decrease with lowering of temperature. Such typical behaviour is reminiscent of weak ferromagnetism (FM) associated with magnetic glassiness or cluster glass (CG) behaviour [44, 45].

Further, insight into the magnetic glassiness of cobalt substituted BFO are obtained from the thermoremnant magnetization (TRM) measurement. For this purpose, two samples are chosen: least Co substituted (BFCO0.5) and highest Co substituted (BFCO2). As the parent BFO does not show indications of magnetic glassiness, it is not considered for this analysis. In this measurement, the sample is first cooled from 320 K to the desired temperature under a magnetic field of 500 Oe. After a wait of 5 minutes ($t_w$) and stabilization of the desired temperature, the magnetic field is switched off and the fall of magnetization with time is recorded. Fig. 4 shows the TRM plots of the two samples at 100 K. Magnetic relaxations in glassy systems are often characterised by a stretched exponential decay function expressed as: $M(t) = M_0 + M_r \exp[-(t/\tau)^{1-p}]$ where, $M_0$ and $M_r$ represents the magnetization of spin clusters and glassy components. The parameters $\tau$ and $p$ are related to relaxation rate. For $0 < p < 1$, the system is a typical cluster glass or spin glass systems [46]. In the present case, single exponential term was found insufficient and a reasonably good fit of the decay curve was obtained with two exponential decay terms in the function. From the fit, the respective relaxation times $\tau_1$ and $\tau_2$ and the exponents $p_1$ and $p_2$ are obtained. The two relaxation times ($\tau_1$ and $\tau_2$) differ by nearly an order of magnitude and its variation with temperature for the above mentioned two samples are shown in the inset of Fig. 4. The variation of two relaxation times for both the samples shows opposite nature: $\tau_1$ increases and $\tau_2$ decreases, with increasing temperature. One more interesting observation which may be noted is that the variation of the relaxation times shows abrupt change near the $T_N$ (~150 K for BFCO2 and ~250 K for BFCO0.5). However, more number of data points in the plot may be needed to make conclusive remark on this aspect. From fitting, it is found that the values of $p_1$ lies between 0.1 - 0.2 whereas the values of $p_2$ are found to be varying from 0.4 (low temperature) to 0.7 (at 250 K). This is an indication of incoherent relaxations for the moments present. The distribution of relaxation rate appears due to magnetic glassiness, which has also been witnessed in the bifurcation of ZFC-FC plot.

Fig. 5 (a-c) shows the isothermal M-H plot of BFO, BFCO1 and BFCO2 up to a field of ±15 kOe, the upper inset shows the first quadrant of M-H data up to 90 kOe. For BFO, a linear behaviour of magnetization versus magnetic field for all the recorded temperatures is observed which goes well with its AFM/PM characteristics as discussed earlier. On the other hand, for the cobalt substituted samples, opening of hysteresis loop hints the onset of ferromagnetic interactions apart from other interactions. Comparison of room-temperature remnant magnetization ($M_r$) and coercive field ($H_c$) for BFCO0.5 and BFCO2 shows a significant enhancement in $M_r$ from 0.0102 emu/g to 0.1401 emu/g and in $H_c$ from 226 Oe to 296 Oe. Moreover, the magnetization (which does not saturate even for H = 90 kOe) values at 90 kOe increases with cobalt content in the sample. Further, for BFCO samples, the value of magnetization at 320 K is in all cases higher than those at 5 K; which is quite unusual in the traditional magnetic materials. A similar feature was also reported in cobalt substituted bismuth ferrite i.e. $Bi(Fe_{1-x}Co_x)O_3$ with $0 \leq x \leq 0.02$ [44]. All these features are typical characteristics of coexisting AFM/PM and weak ferromagnetism (FM), where the later contribution increases with cobalt content in the sample.

In order to strengthen the aforementioned indications of the growing FM interactions in addition to AFM/PM, we have fitted our M-H results to the following expression consisting of FM and AFM/PM parts [47],

$$M(H) = \left[2\frac{M_{FM}^S}{\pi}\tan^{-1}\left\{\left(\frac{H \pm H_{ci}}{H_{ci}}\right)\tan\left(\frac{\pi M_{FM}^R}{2M_{FM}^S}\right)\right\}\right] + \chi H \qquad (1)$$

where, $M_{FM}^S$ is FM saturation magnetization, $M_{FM}^R$ is remnant magnetization, $H_{ci}$ is intrinsic coercivity, $\chi$ is magnetic susceptibility for AFM/PM part, M is the observed magnetization, and H is the applied magnetic field. Here, the first term in the expression represents a ferromagnetic component and the second term is a linear component representing the AFM/PM contribution. The experimental data are fitted via eq.(1) and are shown in Fig. 6, demonstrating a reasonably good fit. The values of the parameters obtained from the fit are tabulated in table 1. The AFM/PM susceptibility $\chi$ increases monotonically from $1.951 \times 10^{-5}$ emu/g·Oe to $2.817 \times 10^{-5}$ emu/g·Oe, as x increases from 0.005 to 0.02, respectively. One aspect of this increase in the present scenario may be considered as the decrease in frustration parameter $f = |\theta_{CW}|/T_N$. Increase in $\chi$ is equivalent to decrease in $1/\chi$ and consequently, the value of intercept in the plot of $1/\chi$ vs T, will be less negative.

Inset of Fig. 5 (b, c) shows temperature variation of coercivity ($H_C$) of BFCO1 and BFCO2 samples. The coercive field ($H_C$) was obtained assuming an average value between positive and negative branches of the M-H loops. In the mean field approach [48], temperature dependence of coercivity follows the relation, $H_C(T) = H_C(0)\{1-(T/T_C)^{1/2}\}$ where, $H_C(0)$ corresponds to the coercivity at 0 K and $T_C$ is the Curie temperature [49]. The coercivity plot in inset of Fig. 5 (b, c) are fitted via this mean field relation and the values of the parameters $H_C(0)$ and $T_C$ obtained are tabulated in table 1. The value of $T_C$ for x = 0.005 is 432 K and the value gradually decreases to 405 K as x increases to 0.02. This observation in association with the susceptibility ($\chi$) in eq. (1), suggests that for small x, nano-sized FM regions are nucleated. These nano-sized FM regions grow in strength and probably in size as well, as the value of x increases to 0.02. However, even at x = 0.02, the size of these FM regions are too small for establishment of any long range interactions, thus leading to cluster-glass behaviour.

The analysis of the first two sections suggests that the substitution of cobalt for Fe-site, is equally probable for the tetrahedral as well as octahedral sites. According to Fig. 1 (a), equi-moment five $Fe^{3+}$ ions of the pentagon: three at tetrahedral sites ($\alpha 1, \alpha 2, \alpha 3$) and two at octahedral sites ($\beta 1, \beta 2$), leads to the observed pentagon frustration. If any of these $Fe^{3+}$ ions ($5.92\,\mu_B$) are replaced with less magnetic $Co^{3+}$ ($4.9\,\mu_B$), the frustration will be lifted and a net magnetic moment (~$1\,\mu_B$) will appear in the pentagon. For instance, if we consider the substituted $Co^{3+}$ to be occupying $\alpha 1$ position, then the magnetic ordering between $\alpha 1$ & $\alpha 2$ (which is AFM) will not be compensated and a non-zero uncompensated magnetic moment at $\alpha 2$ site will give rise to a net magnetization of higher value. As all the five magnetic ions in the crystal lattice (Fig. 1 (a)) are coupled together, they act like a group/cluster. At smaller value of 'x', their probability of being surrounded by another substituted pentagon is less, thus each substituted pentagon behaves like isolated FM cluster. With increase in 'x', more and more pentagons are substituted by $Co^{3+}$ and the chances that a substituted pentagon will be surrounded by another similar one, is more likely. The interaction between two such neighbouring clusters initiates the long-range FM ordering.

**C. Temperature dependant dielectric study:**

The temperature variation of dielectric permittivity (ε') at several frequencies for the samples BFO, BFCO1 and BFCO2 are shown in Fig. 7 (a-c). The insets of the respective plots show the tan loss variation in the same temperature window. The dielectric plot for BFO may be divided in two regions: the low temperature almost linearly rising dispersion-less region and the dispersedly rising region for temperatures above 180 K. At 250 K, which also happens to be $T_N$, the data is slightly oppressed (shown with dotted circle). Consequently, in the tan loss plot also the dispersion opens up at temperature near $T_N$ region and persists till 150 K as the temperature is lowered. This is an indication of coupling of the dielectric relaxation to the magnetic order parameters. A similar behaviour of ε' is seen in the case of BFCO1 and BFCO2, with slightly higher dispersion in the second region. Here too, a slight oppression of data near $T_N$ (shown with an arrow) is observed. The tan loss plots for these two samples are quite different from the parent BFO. With increasing temperature, initially the data remain dispersion-less and from $T_N$ onwards, a constant dispersion plot continues till room temperature. Near room temperature, the tan loss shows a peak which is evident for low frequency data but high frequency peaks falls outside the temperature window and are not observed. With increasing cobalt content, the tan loss peak moves to lower temperatures, thus revealing at least four frequency peaks for BFCO2 whereas only one for BFCO1.

The ε' plot suggest similarity in the dielectric relaxation of the three samples (BFO, BFCO1 and BFCO2) in the form of oppression/hump at $T_N$. On the other hand, the tan loss plot suggests different nature of loss for cobalt substituted samples compared to the parent BFO.

However, one interesting manifestation evident from the dielectric measurement is the plausible ME coupling in the parent BFO as well as cobalt substituted BFCO1 and BFCO2. It is important to note that, the magnetodielectric effect could also appear due to magnetoresistance effect and cannot be related solely to ME coupling [50]. So, it is essential to validate whether the oppression/hump observed in ε' is of capacitive (dielectric) origin or resistive (leakage) origin. As a consequence, we have performed the temperature dependant DC resistance measurement both in zero field as well as 1.3 T magnetic field as illustrated in Fig. 7 (d) for BFCO2 sample. The data is shown from 200 - 305 K, as the value of resistance < 200 K is too high to be measured. Subsequently, we have calculated its magnetoresistance (MR%), defined as $MR\% = [(R(H) - R(0))/R(0)] \times 100$, as shown in the inset of Fig. 7 (d). No significant MR% is seen in BFCO2 which proves that the observed magnetic anomaly in dielectric permittivity (ε") is dominantly of capacitive (dielectric) origin and thus, can be related to true ME coupling. This further proved the understanding of the so-called magnetodielectric effect observed in parent as well as substituted samples. It is to note that, similar features are also seen for BFO and other substituted samples (not shown here).

**D. Magnetodielectric studies:**

Strong enough indications of ME coupling prompted us to investigate the magneto-dielectricity (MD) and magneto-loss (ML) in these samples. The MD and ML are expressed as [51, 52]:

$$MD = \frac{\varepsilon(H) - \varepsilon(0)}{\varepsilon(0)} \times 100 \tag{2}$$

$$ML = \frac{\tan\delta(H) - \tan\delta(0)}{\tan\delta(0)} \times 100 \tag{3}$$

Fig. 8 (a) shows the frequency dependence of MD measured at room temperature under H = 1.3 T. All the three sample BFO, BFCO1 and BFCO2 exhibit negative MD effect, in the measured frequency range at room temperature. The MD for BFO shows maximum value of ~ - 6% at 100 Hz and approaches to zero with frequency increase, non-linearly in the log scale of frequency. Contrary to this, the BFCO1 and BFCO2 samples show nil or very low MD at 100 Hz and peaks at ~ 3 – 4 kHz and again approaches zero or lowest value. The MD plot is found to obey the power law of the form, $MD \propto \exp(\omega)^n$. For BFO sample, the exponent $n$ ~ -0.5 (>5 kHz) and n ~ -0.2 (<5 kHz) shows a reasonable good fit while for cobalt substituted BFCO1 and BFCO2, the low frequency behaviour changed drastically, and the power law fitting suggests the exponent to be $n$ ~ 0.2. This value of $n$ ~ 0.2 is quite close to the magnetization exponent '$1-p_2$ ~ 0.3' of the thermoremanent measurement. This shows a strong correlation of dielectric with the magnetization i.e., a plausible magnetoelectric (ME) coupling. The high frequency data again fits for $n \sim -0.5$, as in parent BFO. Thus, the MD result suggests that two contributions are mainly responsible for the observed MD in these samples: a low frequency process and a high frequency process. The high frequency process, which may also be regarded as the intrinsic contribution, remains same in parent BFO as well as substituted BFCO1 and BFCO2. On the other hand, the low frequency process, which is vulnerable to dc conductivity, leakage currents, space-charge polarization etc., makes a drastic change for cobalt substituted BFCO1 and BFCO2 samples, compared to parent BFO. Inset of Fig. 8 (a) shows ML% versus frequency of BFO, BFCO1 and BFCO2. The ML for BFO shows maximum value of $\sim -7\%$ at 100 Hz and approaches to zero, almost linearly (in log scale) in the high frequency region. While a negative ML ($\sim 2 - 3\%$) is seen for BFCO1 and BFCO2 samples at 100 Hz and then attains a positive value in the mid frequency range (0.3 - 5 kHz) and finally decreases to least negative ML with increase in frequency. A positive ML% in the region 300 Hz - 5 kHz may have resulted due to increase in conductivity in the modified samples due to an external field of 1.3 T. Further, to ascertain the true nature of polarisation, imaginary part of dielectric constant ($\varepsilon''$) versus frequency plot was analysed carefully. Fig. 8 (b) shows frequency dependence $\varepsilon''$ at room temperature for BFCO2 sample. In general, if $\varepsilon''$ shows a symmetric peak at some non-zero $\omega$ and $\varepsilon'' \to 0$ as $\omega \to 0$, this signifies an intrinsic Debye type polarisation, while in Maxwell-Wagner (MW) type polarisation $\varepsilon''$ diverges at low frequency [53]. In an attempt to fit the data with intrinsic dipolar relaxation mechanism, we observe that the fitting gets deviated at low frequency (<5 kHz) while the fitting is good at higher frequencies (>5 kHz), as shown in the inset of Fig. 8 (b). This feature indicates the presence of MW type mechanism at low frequency. Thereafter, the $\varepsilon''$ versus frequency data is fitted to a generalised Haverliak-Negami (HN) relaxation expression in addition to a finite conductivity contribution which is expressed as: [54]

$$\varepsilon^*(\omega) - \varepsilon_\infty = \frac{\varepsilon_s - \varepsilon_\infty}{[1 + (i\omega\tau)^\alpha]^\beta} - i\left[\frac{\sigma_{dc}}{\varepsilon_0 \omega}\right]^n \qquad (4)$$

where, $\varepsilon_s$ - $\varepsilon_\infty$ is the static dielectric constant of the relaxation present in the system, $\tau$ is the corresponding relaxation time, $i = \sqrt{-1}$, $\sigma_{dc}$ is dc conductivity, $\varepsilon_0$ is the permittivity of free space, $\omega$ (=2$\pi f$) is the angular frequency and exponent $n$ lies between 0 and 1. The parameters $\alpha$ and $\beta$ denotes asymmetry and broadness of the corresponding relaxation spectra. In eq. 4, the first term represents the polar nature of relaxation (such as Debye) and the second term is due to the contribution from finite conductivity (such as MW). Thus, our analysis shows an involvement of finite conductivity contribution of the form $(1/\omega)^n$ at low frequency (< 5 kHz) which is supposed to have arisen due to the extrinsic effect whereas, the high frequency region (> 5 kHz) is found to be showing intrinsic contribution. As a result, all our magnetodielectric study in the intrinsic region are of polar nature.

The temperature variation of MD% and ML% for parent BFO and substituted BFCO2, at a selected frequency say 12 kHz is shown in Fig. 9 (a, b). Here, we find a contrasting behaviour of the MD for the two samples. The BFO shows a negligible MD and ML at low temperatures but as $T_N$ (250 K) approaches, a sudden rise in negative MD and ML is seen. Even after crossing the $T_N$, the rise continues and at 300 K it becomes ~ -0.3% (MD) and ~ -3% (ML). Contrastingly, the BFCO2 shows a nearly constant negative value of –2.4% at low temperatures and as its $T_N$ (152 K) approaches, the magnitude shows a linear increase with increasing temperature and reaches room temperature with value -0.9% (MD). This marks about twice enhancement of MD% at 300 K in BFCO2 as compared to that of parent sample (BFO). An extrapolation of the linear region above $T_N$, suggests that MD may become zero for T ~ 365 K, which is very close to the $T_{irr}$ (~ 370 K) obtained from χ versus T plot. On the other hand, ML of BFCO2 shows a higher positive value of ~ 80% in the temperature range of 150 - 180 K which is in the vicinity of $T_N$ (152 K). Such a high value of ML around $T_N$ indicates that the ME coupling in cobalt substituted samples are different in nature compared to the parent BFO. A similar feature is also seen for BFCO1 sample (not shown here).

Field dependence of MD% is shown in Fig. 10. The data are collected at 50 kHz in order to nullify the effect of interfacial space-charge artifacts [55]. Negative MD% is seen for all the samples with value of ~ -0.03% (BFO), ~ -0.05% (BFCO1) and ~ -0.1% (BFCO2) at applied field of 1.3 T. The field dependence of MD% is found to obey power law of the form $MD\% \propto (H)^m$. The values of 'm' are found to lie in the range 0.4 - 0.5 which is very close to the values of 'n' in the frequency dependence of MD in the intrinsic region, thus revealing a similar dependence of MD on frequency as well as magnetic field.

It thus seems that the MD induced in substituted samples are dominated by the spin disorderness, whereas that in BFO is mainly due to freedom gained against the magnetic ordering above the magnetic ordering temperature ($T_N$). Here, the role of antiferromagnetic ordering seems to suppress the MD. That is why almost negligible MD is seen below $T_N$ in BFO. In BFCO2, further increase in MD is also stopped at $T_N$ and the MD value gets arrested there itself.

## IV. CONCLUSION

In conclusion, structural, magnetic, magnetodielectric behaviour of $Bi_2(Fe_{1-x}Co_x)_4O_9$ ($0 \leq x \leq 0.02$) samples were investigated. Irreversibility around 370 K in ZFC-FC curves of modified samples revealed weak FM associated with magnetic glassiness or CG behaviour, successively confirmed from TRM data. The AFM transition ($T_N$) is found to decrease with increase in x content. Substitution of less magnetic $Co^{3+}$ ($4.9\,\mu_B$) ions at $Fe^{3+}$ site ($5.92\,\mu_B$) generated a non-zero uncompensated magnetic moment leading to the rise in net magnetisation, initiating long-range FM ordering. At room temperature, a significant improvement of values in $M_r$ from 0.0102 emu/g (BFCO0.5) to 0.1401 emu/g (BFCO2) and in $H_c$ from 226 Oe (BFCO0.5) to 296 Oe (BFCO2) was noticed. Moreover, the presence of a concomitant anomaly in the temperature dependant dielectric data around $T_N$ indicated the evidence of magneto(di)electric coupling in all the samples. Enhanced MD effect along with the existence of magnetic glassiness in substituted samples is governed by spin disorderness. A significant enhancement of two times in MD effect is obtained for $x = 0.02$ as compared to $x = 0$. Thus, we hope the above features would incline to further scientific research and make it a viable candidate for diverse multifunctional device applications.

**Acknowledgement**

AKS acknowledges Board of Research in Nuclear Science (BRNS), Mumbai (Sanction No: 2012/37P/40/BRNS/2145), UGC-DAE-CSR Mumbai (Sanction No: CRS-M-187, 225) and Department of Science and Technology (DST), New Delhi (Sanction No: SR/FTP/PS-187/2011) for funding. Lastly, SRM is thankful to BRNS for the financial assistance.

**FIGURE CAPTIONS**

FIG. 1 (Colour Online): (a) An unit cell of $Bi_2Fe_4O_9$ showing projection of $Fe^{3+}$ ions in the a-b plane where α1, α2, α3 and β1, β2 denotes tetrahedral (Fe(T)) and octahedral (Fe(O)) coordinated Fe ions, respectively. Collectively, the arrangement of α1, α2, α3, β1 & β2 depicts a unique pentagon frustration. $J_3$, $J_4$ and $J_5$ are exchange interactions between β1- α1 (β2- α3), α1-α2 and β2- α2 (β1- α3), respectively. (b-f) Rietveld refinement of room-temperature X-ray diffraction patterns of BFO, BFCO0.5, BFCO1, BFCO1.5 and BFCO2, respectively.

FIG. 2 (Colour Online): XPS spectrum of the Fe-2p spectra of (a) BFO and (b) BFCO2.

FIG. 3 (Colour Online): (a-c) Temperature-dependence of ZFC and FC magnetizations of BFO, BFCO1 and BFCO2 samples under 1 kOe. Insets (a-c) shows the dχ/dT plots indicating the magnetic transitions ($T_N$). Lower inset of Fig. 3(a) shows 1/χ versus temperature plot of BFO demonstrating the frustration parameter (*f*) and the red solid line is the Curie-Weiss fit.

FIG.4 (Colour Online): Time-dependant thermoremanent magnetization (TRM) at 100 K for BFCO0.5 and BFCO2 samples. The solid line is a fitting curve. Inset shows the plot of temperature variation of relaxation times ($τ_1$ and $τ_2$) of BFCO0.5 and BFCO2.

FIG. 5 (Colour Online): The isothermal M-H plots of (a) BFO, (b) BFCO1 and (c) BFCO2 samples at 5 K and 320 K, in the range ±15 kOe. Upper insets (a-c) shows the first quadrant plot of M-H curve up to 90 kOe. Lower insets (a-c) shows coercive field ($H_C$) vs temperature with the theoretical fit.

FIG. 6 (Colour Online): Experimental magnetization data of BFCO2 sample with the fitted (solid line) and theoretically simulated AFM/PM and FM part at 320 K, in the range ±90 kOe. Inset shows the expanded view, in the range ±9.5 kOe.

FIG. 7 (Colour Online): Dielectric permittivity (ε') and tan loss (insets) versus temperature plots of (a) BFO, (b) BFCO1 and (c) BFCO2 at several frequencies. Dotted circle (BFO) and arrows (BFCO1 and BFCO2) reflect the onset of magnetic transition ($T_N$). (d) Temperature dependence of DC resistance measurement at 0T and 1.3T of BFCO2 and inset shows temperature dependant MR%.

FIG. 8 (Colour Online): (a) Room temperature frequency variation of MD% and ML% (inset) of BFO, BFCO1 and BFCO2 samples under H = 1.3T. Dotted lines show the power law fitting of the form $MD \propto \exp(\omega)^n$. (b) Frequency dependence of ε" of BFCO2 at room temperature fitted to eq. 4 and inset shows the poor fitting when data is fitted only to Haverliak-Negami relaxation mechanism.

FIG. 9 (Colour Online): Temperature dependent (a) MD% and (b) ML% of BFO and BFCO2 at 12 kHz. The solid lines are guide to the eye.

FIG. 10 (Colour Online): Field dependant MD effect of BFO, BFCO1 and BFCO2 samples at 50 kHz, 320 K. The solid lines are fitted using power law of the form $MD \propto (H)^m$.

**Table:**

Table 1: A compilation of lattice parameters, crystallite size, micro-strain and other fitted parameters of $Bi_2(Fe_{1-x}Co_x)_4O_9$ ($0 \leq x \leq 0.02$) samples.

| Samples → | BFO | BFCO0.5 | BFCO1 | BFCO1.5 | BFCO2 |
|---|---|---|---|---|---|
| Lattice parameters | | | | | |
| a /pm | 797.350(2) | 797.318(5) | 796.806(4) | 795.993(2) | 795.262(3) |
| b /pm | 844.142(4) | 844.113(2) | 843.627(6) | 843.008(6) | 842.397(6) |
| c /pm | 600.277(5) | 600.271(2) | 599.891(2) | 599.028(5) | 599.002(2) |
| Volume /$10^6$ $pm^3$ | 404.032(4) | 403.998(7) | 403.250(1) | 401.964(5) | 401.287(3) |
| Reduced $\chi^2_{XRD}$ | 3.66 | 2.94 | 2.78 | 3.02 | 2.96 |
| Crystallite size /nm | 247(9) | 138(2) | 127(4) | 121(2) | 118(3) |
| Micro-strain /$10^{-3}$ % | 2.20(7) | 7.13(5) | 7.31(2) | 7.98(5) | 8.36(7) |
| Fitting parameters from Fig. 5 and Fig. 6 | | | | | |
| $M^R_{FM}$ /emu/g at 320 K | - | 0.006(1) | 0.018(2) | 0.025(2) | 0.052(1) |
| $M^S_{FM}$ /emu/g at 320 K | - | 1.21(2) | 1.53(3) | 1.62(5) | 1.73(2) |
| $\chi^{PM/AFM}$ /$10^{-5}$ emu/g·Oe | - | 1.951(16) | 2.244(22) | 2.573(27) | 2.817(29) |
| $T_c$ /K | - | 432(4) | 426(8) | 413(5) | 405(7) |
| $|H_C|$ /Oe | - | 229(4) | 260(9) | 269(6) | 299(10) |

Table 2: Interatomic metal-oxygen distances [pm] of BFO and BFCO2 samples obtained from Rietveld refined room-temperature XRD data.

| Samples → | BFO | BFCO2 |
|---|---|---|
| Bi-O11 | 244.51(5) | 242.52(9) |
| Bi-O11 | 214.32(7) | 219.20(4) |
| Bi-O2 (×2) | 216.89(14) | 223.58(16) |
| Bi-O2 (×2) | 305.53(5) | 303.83(4) |
| Fe1-O11 (×2) | 207.62(7) | 211.29(8) |
| Fe1-O12 (×2) | 196.16(9) | 186.52(3) |
| Fe1-O2 (×2) | 202.74(4) | 208.91(6) |
| Fe2-O12 | 179.13(8) | 180.56(11) |
| Fe2-O2 (×2) | 193.57(15) | 190.18(9) |
| Fe2-O3 | 180.86(6) | 185.82(13) |

Table 3: Positional parameters of BFO and BFCO2 samples obtained from Rietveld refined room-temperature X-ray diffraction data.

| Atoms | Wyckoff | Variables | Rietveld refined values | |
|---|---|---|---|---|
| | | | BFO | BFCO2 |
| Bi | 4g | x | 0.32304(19) | 0.33050(12) |
| | | y | 0.17179(14) | 0.16752(16) |
| | | z | 0 | 0 |
| Fe1(O)/Co1 | 4e | x | 0 | 0 |
| | | y | 0 | 0 |
| | | z | 0.25135(17) | 0.26310(28) |
| | | Occu. Co* | - | 0.056(8) |
| Fe2(T)/Co2 | 4h | x | 0.14817(16) | 0.15042(13) |
| | | y | 0.32214(11) | 0.33529(8) |
| | | z | 0.5 | 0.5 |
| | | Occu. Co* | - | 0.082(6) |
| O11 | 4g | x | 0.33949(21) | 0.33991(18) |
| | | y | 0.42708(17) | 0.42877(12) |
| | | z | 0 | 0 |
| O12 | 4h | x | 0.36361(19) | 0.37147(15) |
| | | y | 0.42694(21) | 0.42223(17) |
| | | z | 0.5 | 0.5 |
| O2 | 8i | x | 0.13115(14) | 0.13412(16) |
| | | y | 0.20689(17) | 0.21598(24) |
| | | z | 0.25183(16) | 0.26068(12) |
| O3 | 2d | x | 0 | 0 |
| | | y | 0.5 | 0.5 |
| | | z | 0.5 | 0.5 |

* $Occu.$ (Fe) = 1 – $Occu.$ (Co)

**FIGURES**

**FIG. 1**

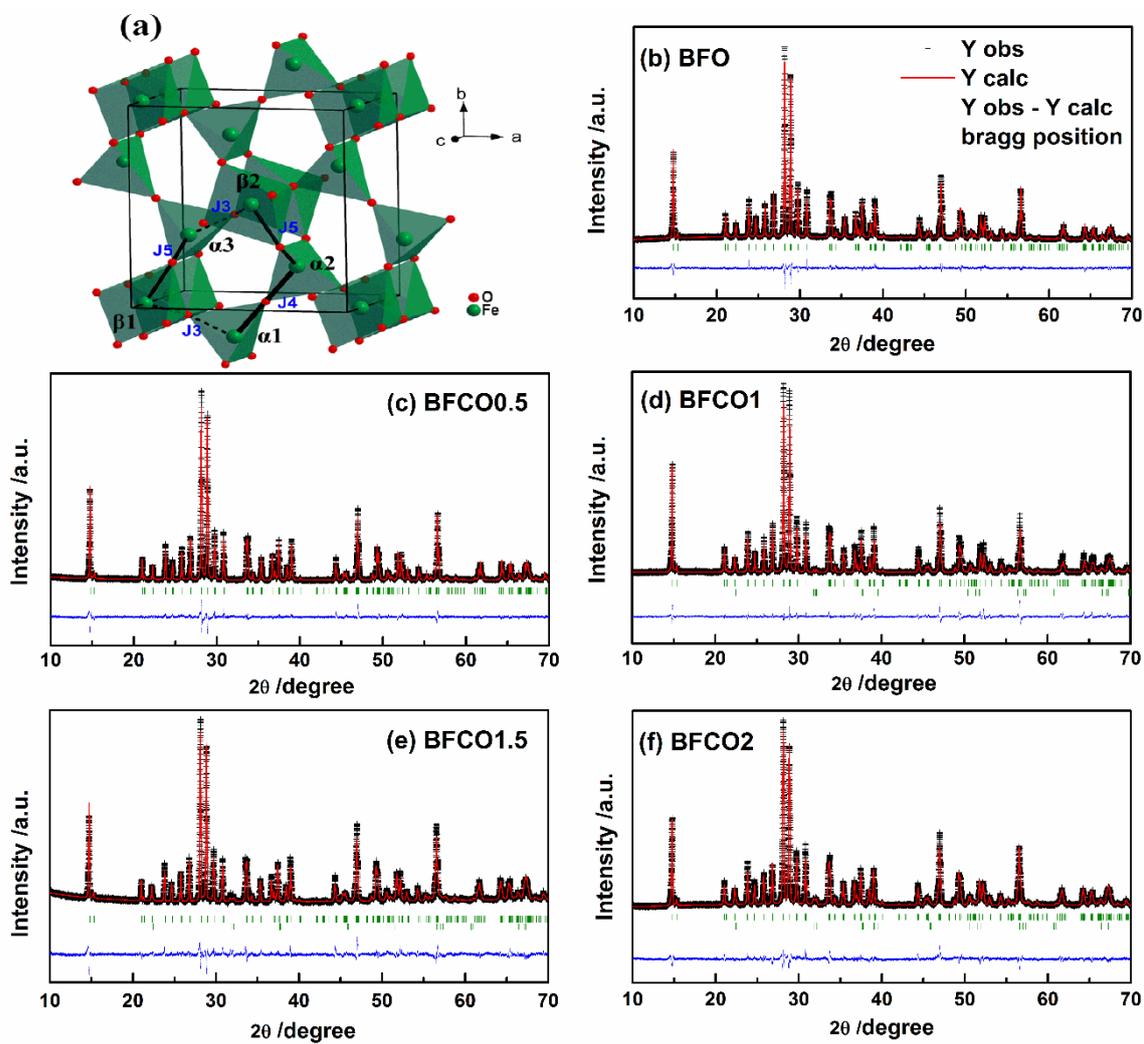

**FIG. 2**

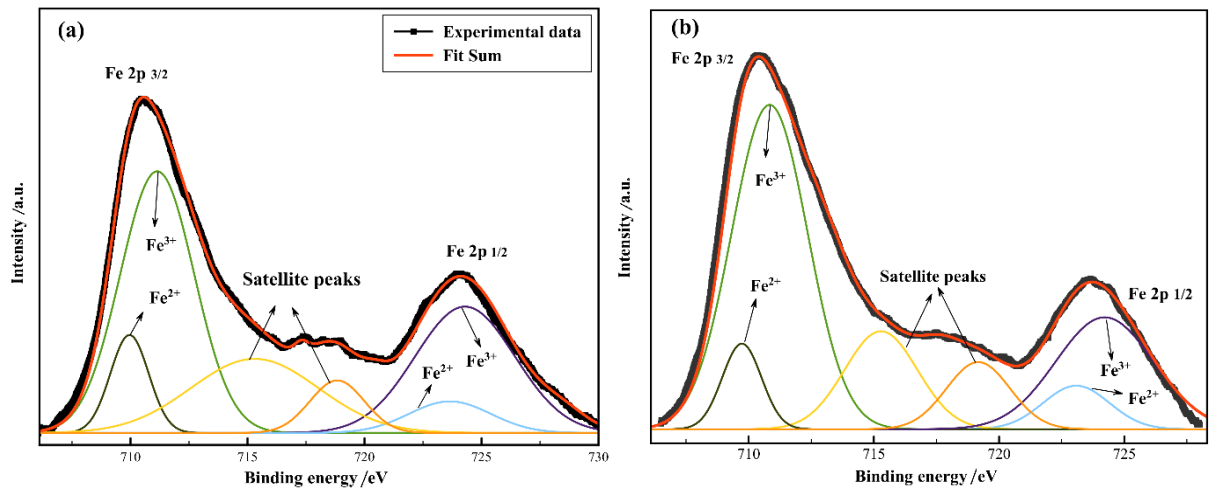

**FIG. 3**

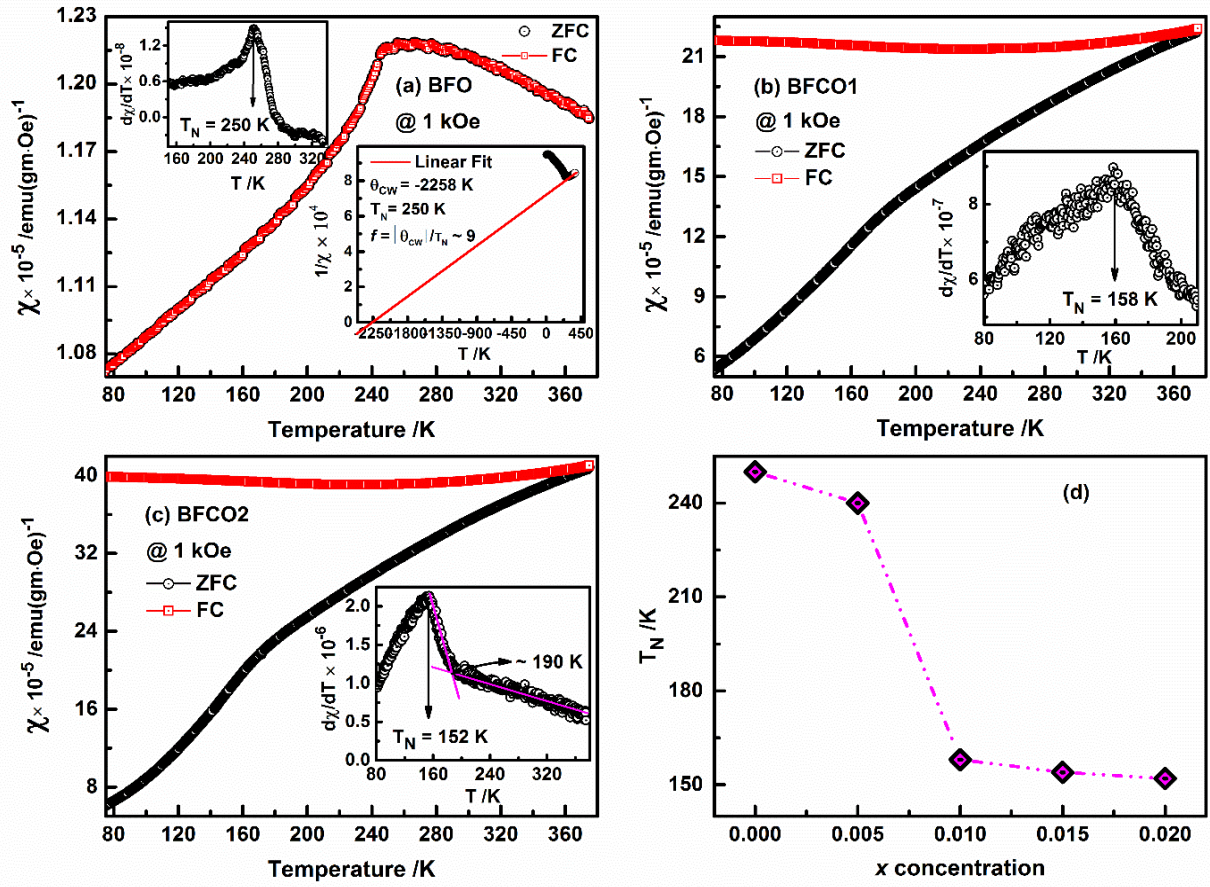

**FIG. 4**

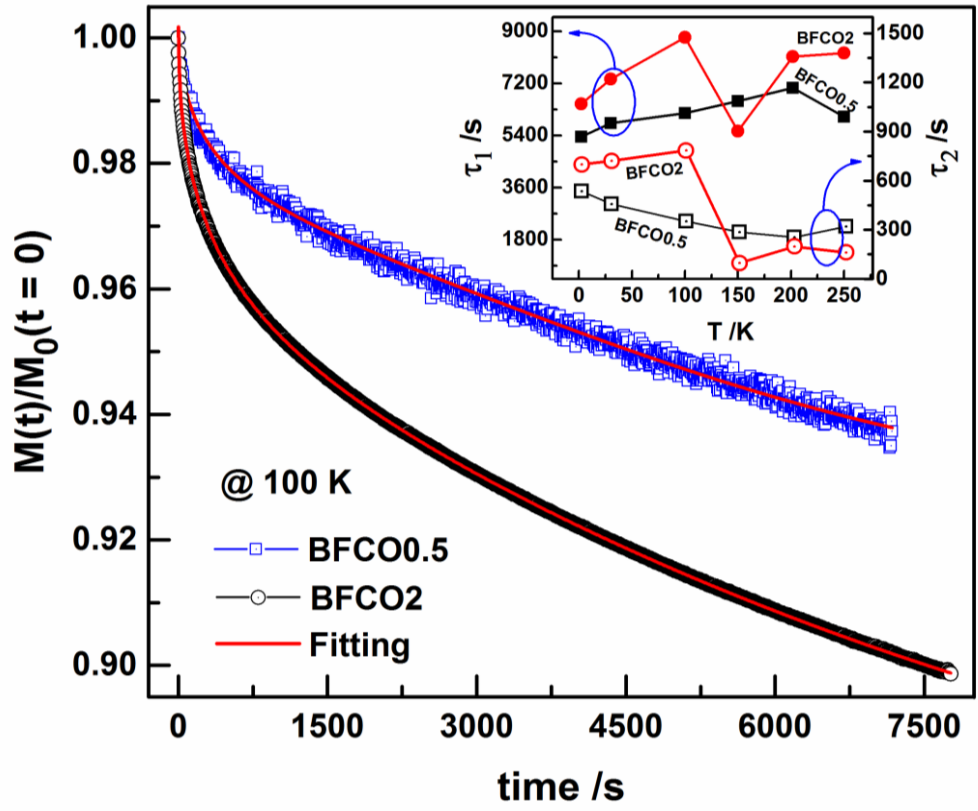

**FIG. 5**

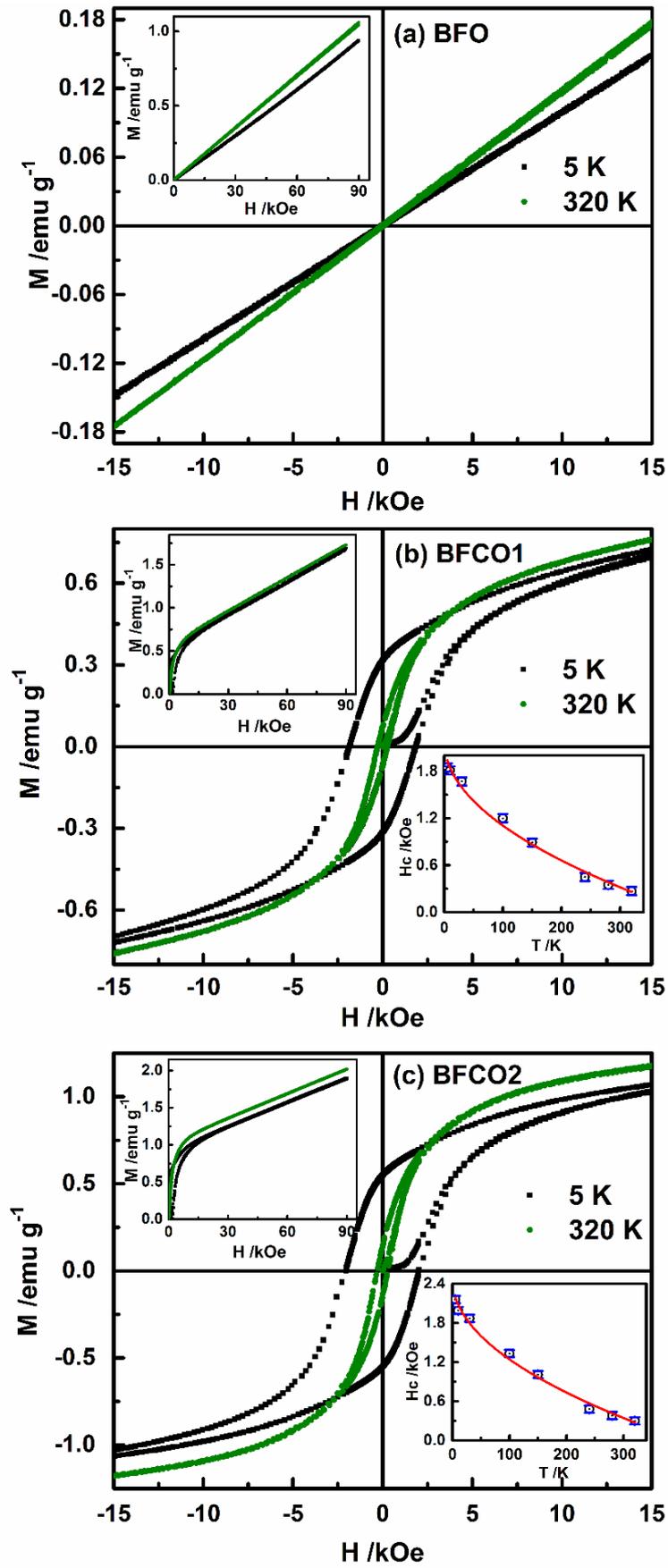

**FIG. 6**

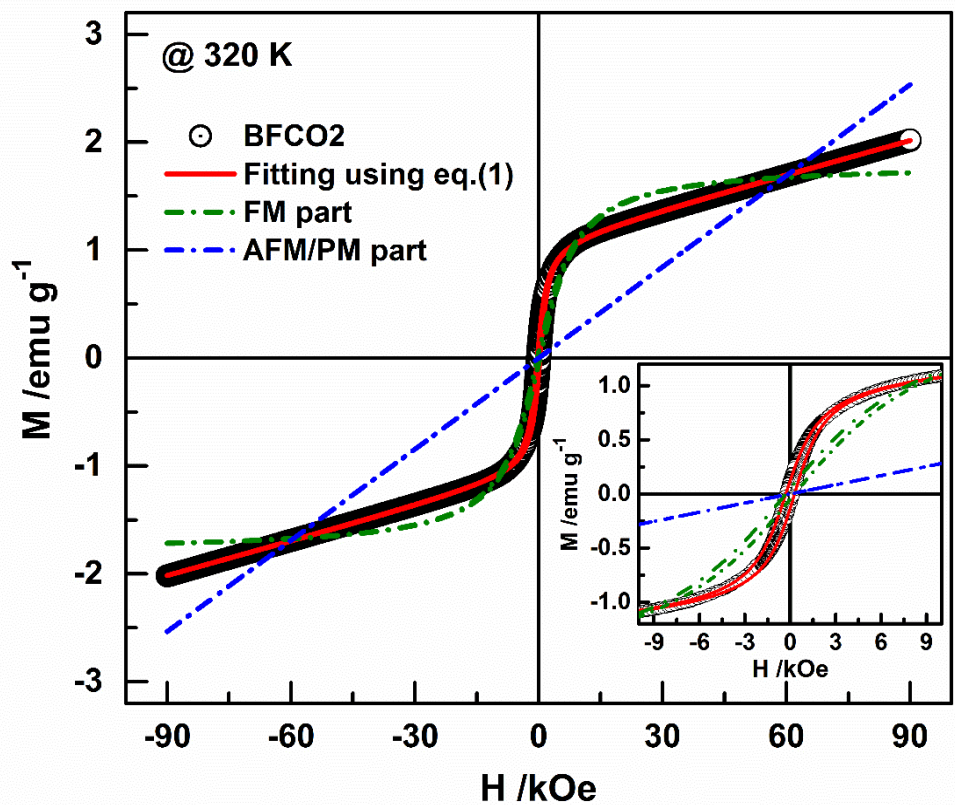

**FIG. 7**

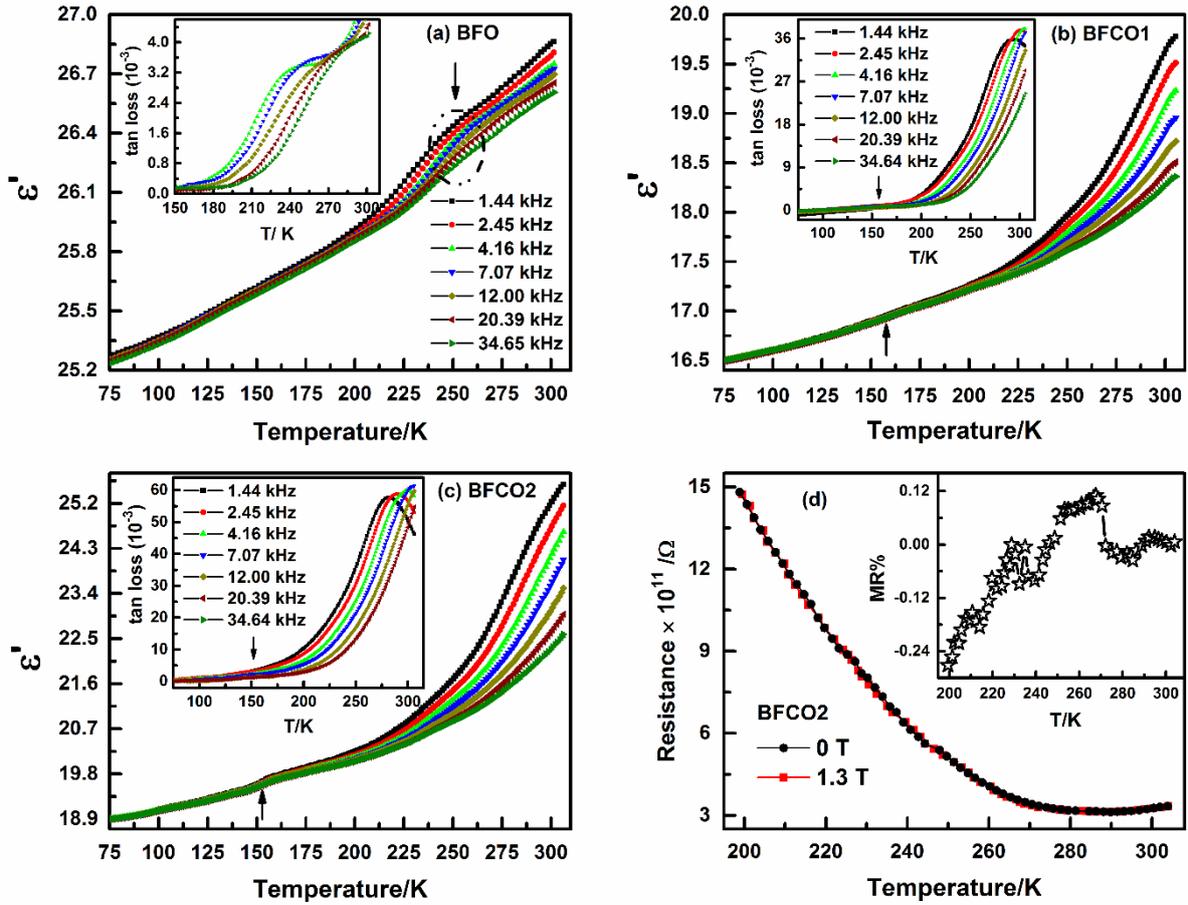

**FIG. 8**

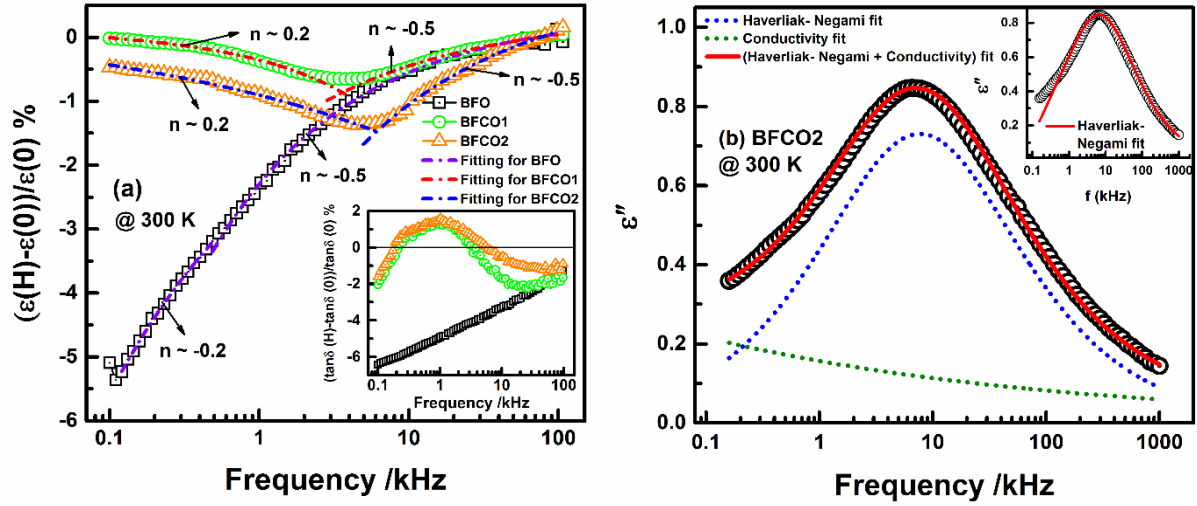

**FIG. 9**

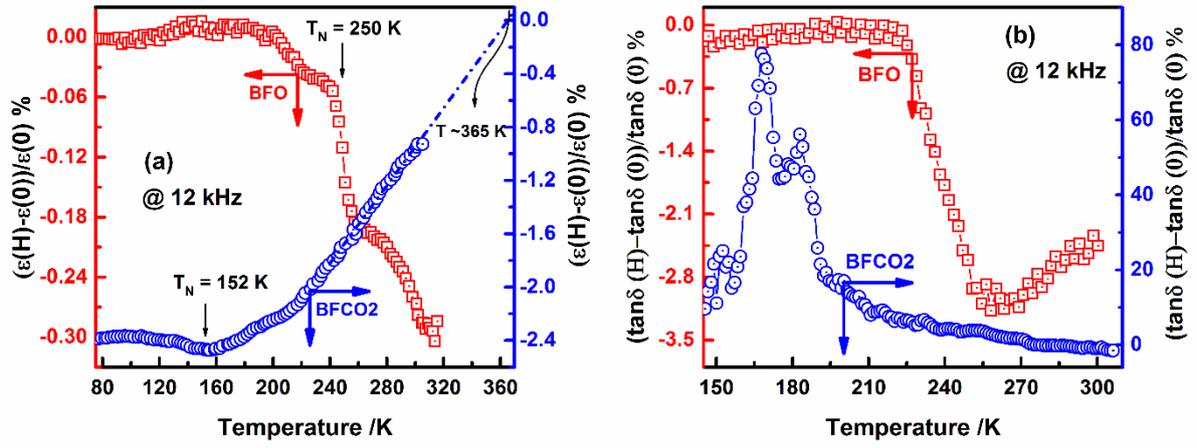

**FIG. 10**

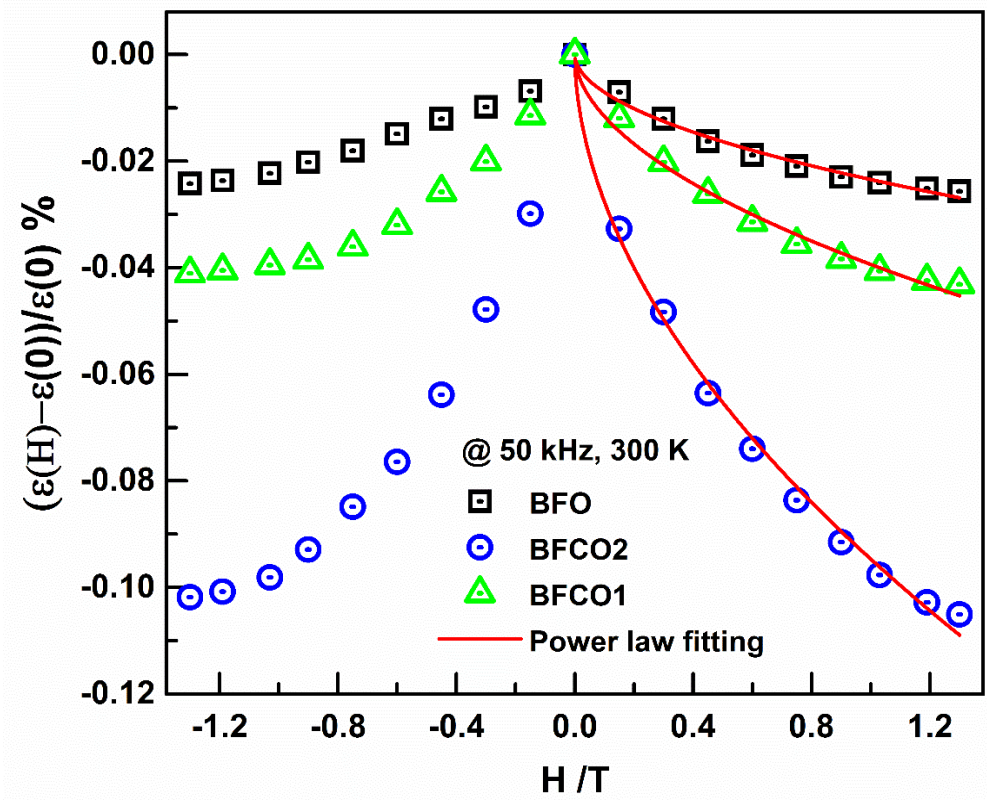